\documentclass{elsarticle}

\usepackage{lineno,hyperref}

\usepackage[utf8]{inputenc}
\usepackage{comment}

\usepackage{xcolor}
\usepackage{ulem}

\usepackage[utf8]{inputenc}
\usepackage{amsmath,amssymb,amsthm,amssymb}
\usepackage{datetime}
\usepackage{graphicx}
\usepackage{listings}
\usepackage{xspace}
\usepackage{natbib}

\newcommand{\musicsim}{\lstinline[columns=fixed]{musicsim}\xspace}

\journal{Nuclear Instruments and Methods in Physics Research Section A: Accelerators, Spectrometers, Detectors and Associated Equipment}

\date{\today}

\begin{document}



\begin{frontmatter}

\title{Classification of events from $\alpha$-induced reactions in the MUSIC detector via statistical and ML methods}

\author[MCS]{K.~Raghavan}
\author[PHY]{M.~L.~Avila}
\author[MCS]{P.~Balaprakash}
\author[PHY]{H. Jayatissa}
\author[PHY]{D.~Santiago-Gonzalez}
\cortext[mycorrespondingauthor]{Corresponding author}
\ead{dsg@anl.gov}

\address[MCS]{Mathematics and Computer Science Division, Argonne National Laboratory, Lemont, IL 60439, USA}
\address[PHY]{Physics Division, Argonne National Laboratory, Lemont, IL 60439, USA}

\begin{abstract}
The Multi-Sampling Ionization Chamber (MUSIC) detector is typically used to measure nuclear reaction cross sections relevant for nuclear astrophysics, fusion studies, and other applications. From the MUSIC data produced in one experiment scientists carefully extract an order of $10^3$ events of interest from about $10^{9}$ total events, where each event can be represented by an 18-dimensional vector.
However, the standard data classification process is based on expert-driven, manually intensive data analysis techniques that require several months to identify patterns and classify the relevant events from the collected data.
To address this issue, we present a method for the classification of events originating from specific $\alpha$-induced reactions by combining statistical and machine learning methods that require significantly less input from the domain scientist, relative to the standard technique.
We applied the new method to two experimental data sets and compared our results with those obtained  using traditional methods. With few exceptions, the number of events classified by our method agrees within $\pm20\%$ with the results obtained using traditional methods.
With the present method, which is the first of its kind for the MUSIC data, we have established the foundation for the automated extraction of physical events of interest from experiments using the MUSIC detector.

\end{abstract}

\begin{keyword}
AI/ML methods\sep $\alpha$-induced reactions \sep active target\sep experimental nuclear physics
\end{keyword}

\end{frontmatter}


\section{Introduction}

The quest for answers to some of the most fundamental questions in nuclear physics has led to the design and construction of increasingly  complex experimental equipment and facilities. Through  careful analysis of the experimental data generated by detector systems,  nuclear physicists make discoveries and measure properties of atomic nuclei for basic research and diverse applications. The scale of the experiments, detector systems, and collected data demands advances in current analysis techniques in order to efficiently extract the relevant information from such large and complex data sets. At present, expert-driven, manually intensive data analysis techniques require several months to identify patterns and classify the relevant events from the collected data. Artificial intelligence (AI) techniques, in particular machine learning (ML) methods, have the potential to automate, accelerate, and improve the data analysis techniques, thereby reducing the effort to obtain the  physical data, enhancing the scientific throughput, and potentially enabling further discoveries.

From the AI/ML perspective, the analysis of data produced by the detector systems poses a number of unique challenges that need the development of domain-specific AI/ML methods. For example, in the  case of the Multi-Sampling Ionization Chamber (MUSIC) detector data, the number of background events is typically several orders of magnitude more  than the events of interest, rendering out-of-the-box anomaly detection methods ineffective. A few recent examples of domain-specific AI/ML methods can be found in the works of Kuchera et al.~\cite{kuc19} (experiment) and Raghavan et al.~\cite{Rag21} (theory). For a recent review, see \cite{Boe21}.

In this paper we present a method for the classification of data obtained from MUSIC, an active-target detector specializing in cross-section measurements of various nuclear reactions such as fusion~\cite{Carnelli15} and $\alpha$-induced reactions~\cite{AvilaNIM16}.

Our proposed approach comprises four phases: (1) filter easy-to-identify background events; (2) detect and remove background events that are similar to the events of interest; (3) get input from the domain scientists and design a classifier; and (4) detect the strip location at which the event took place. We leverage statistical and machine learning methods in phases 1--3. In addition to experimental data, we  use simulated data in phases 1--3 to assist in the event classification.

We applied our method  to two data sets obtained under different experimental conditions. In the first data set, $\alpha$-induced reactions on Fluorine-17 ($^{17}$F) nuclei were observed and classified with standard analysis techniques. In the second data set, $\alpha$-induced reactions on Oxygen-17 ($^{17}$O) nuclei were studied. The $^{17}$O data was previously classified with  standard techniques, and those results are published in \cite{AvilaNIM16}.

We show that our new method classifies the MUSIC data---that is, it separates relevant events for a particular nuclear reaction from other events in a given MUSIC strip---in a way such that the percent differences with respect to human expert classification are typically within $\pm20\%$.%

This paper is organized as follows. In Sec.~\ref{s:music} we briefly describe the MUSIC detector and its data structure.  In Sec.~\ref{s:exp} we give a brief description of the experiments that produced the data analyzed by the methods presented in this paper. In Sec.~\ref{s:simulation} we present basic information on the simulation of the detector response. In Sec.~\ref{s:methods} we describe our data classification method, which is based on statistical and ML methods. In Sec.~\ref{s:results} and \ref{s:conclusion} we present results and conclusions, respectively.

\section{The MUSIC detector and its data structure}
\label{s:music}

MUSIC is an active-target gas-filled ionization chamber, where the gas serves as both target and detection material. The MUSIC detector has been used to measure excitation functions of heavy-ion fusion reactions~\cite{Carnelli14,Carnelli15} and to measure $(\alpha,p)$ and $(\alpha,n)$ reaction cross sections in inverse kinematics~\cite{AvilaNIM16,Avila16,Talwar18,AvilaCP18}, involving stable and radioactive beams. For the $(\alpha,p)$ and $(\alpha,n)$ reactions, cross sections down to 1~mb have been measured. If the $(\alpha,p)$ and $(\alpha,n)$ channels are both energetically allowed at the energies studied, then the MUSIC detector is able to probe them simultaneously~\cite{Avila16}.  The high detection efficiency of MUSIC (close to 100\%) makes the most of heavy-ion beams with relatively low intensity (less than $10^6$ particles per second). In addition, MUSIC can measure averaged cross-section points at different beam energies with a single incident beam energy. The reason is that the incident beam particles lose energy through interactions with the gas molecules as they move deeper into the detector. 

\begin{figure}
    \centering
        \includegraphics[scale=0.25]{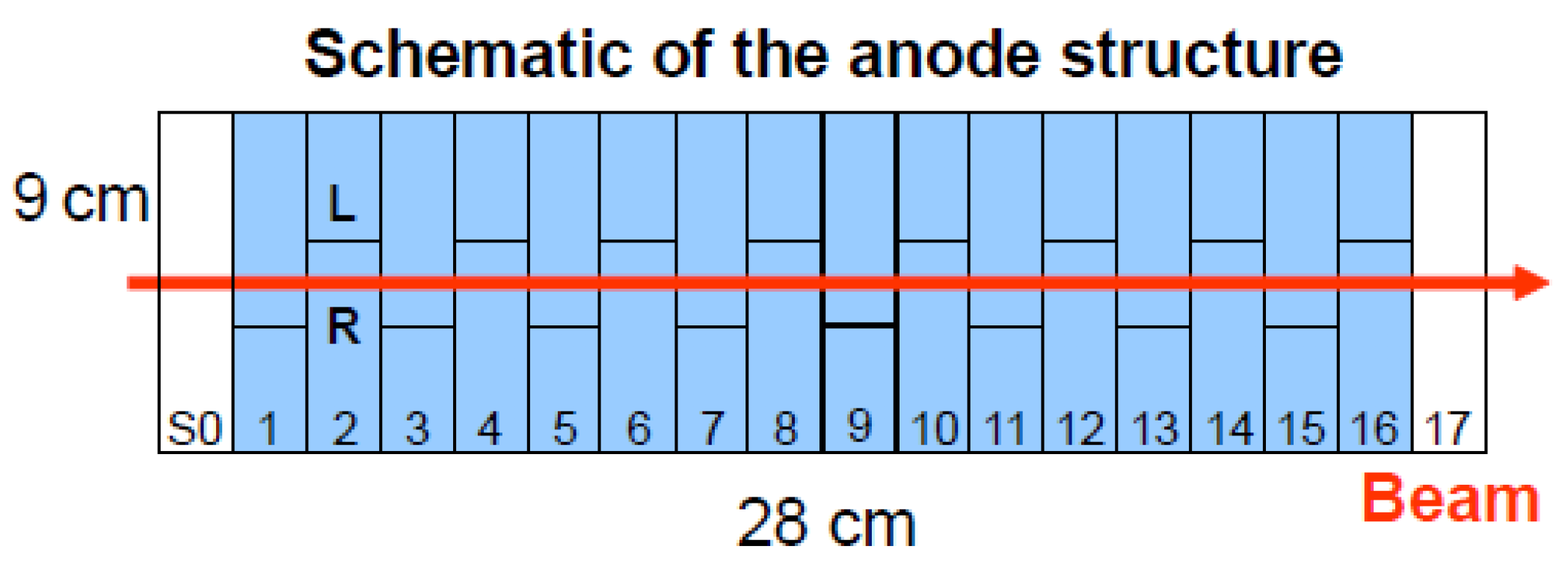}
    \caption{\label{fig:MUSIC} 
    Schematic of the MUSIC detector anode structure. There are 18 strips composed by 34 electronics channels.}
\end{figure}

The anode of MUSIC has 18 segments, with the middle 16 segments subdivided in left and right as shown in Fig.~\ref{fig:MUSIC}. With this segmentation, the total number of electronic channels in the anode is 34. The data from the left and right segmentation has proven useful when studying heavy-ion fusion reactions, since the outgoing particles from elastic scattering events deposit enough energy in both the left and right segments of a single strip (multiplicity 2) and can be clearly separated from fusion events that have multiplicity 1; see~\cite{Carnelli15}.  For $\alpha$-induced reactions, however, separation of elastic scattering events from the  events of interest based on ``multiplicity'' is not as effective since the energy deposited by the elastically scattered helium ions in the gas is such that it is normally comparable to the electronics noise. Therefore, only the sum of the left and right segments was used for the ML analysis of the $\alpha$-induced reactions presented in this paper and not the individual contributions for left and right. This reduces the number of dimensions of an event from 34 to 18. For a more detailed description of the detector and the standard data analysis technique, see \cite{Carnelli14,AvilaNIM16}.

While the data collection of an experiment performed with the MUSIC detector is carried out in a couple of days, it usually takes a human expert  several months to both find the relevant information from the $\sim10^{9}$ recorded events and to classify the $\sim10^3$ events of interest, in order to then extract the reaction cross sections.  
In order to separate the background events from the events of interest, henceforth referred to as \textbf{golden events}, vectors containing the particle's energy loss in each detector segment are studied. The energy loss or energy deposited by different particles in a segment is described by the Bethe formula~\cite{Bet30} that, in the present energy regime, is proportional to the square of the number of protons, $Z$. Therefore, under appropriate experimental conditions, particles in MUSIC can be separated if they have different $Z$ number. To distinguish between different reactions, the energy loss in several MUSIC segments, after the reaction occurs, is summed and plotted against the total energy deposited in the detector---the so-called $\Delta E$--$E$ plots. Depending on the energy and type of reaction, sometimes several permutations of the plots have to be performed in order to optimize the separation of different events. Then, two-dimensional graphical ``cuts'' are used to select promising events, which are later classified by studying the characteristic shape of their energy loss profiles across the segmented anode strips, known as \textit{event traces}.
Event traces are further discussed in Sec.~\ref{s:simulation} (examples are shown in Fig.~\ref{fig:simVal}).

Table~\ref{tab:datastruct} shows basic information about the data structure of an event in the MUSIC detector. Currently, the data from each of the 18 strips of the MUSIC detector can be acquired by using  analog or digital electronics. In the analog case, 10-bit Philips analog to digital converters, controlled by a Computer-Aided Measurement And Control (CAMAC) crate, are used. In the digital case, 14-bit CAEN digitizers, controlled by a Versa Module Eurocard (VME) crate, are used.

\begin{table}[h!]
    \centering
    \begin{tabular}{c|c}
        Dimension & 18 \\
        Raw data type & integer\\
        Raw data range & 0 to 4,096, or 0 to 16,384\\
        Calibrated data type & real (floating point)\\
        Calibrated data range & depends on experimental conditions
    \end{tabular}
    \caption{Data structure of a MUSIC event (point). Raw data range depends on the data acquisition system (10- or 14-bit) used, but it is constant during a single experiment.}
    \label{tab:datastruct}
\end{table}

\section{Experimental data}
\label{s:exp} 

The experimental data shown in this paper was obtained with the MUSIC detector from experiments aimed to probe the $\alpha$-induced reactions $^{17}$F$(\alpha,p)^{20}$Ne and $^{17}$O$(\alpha,n)^{20}$Ne. These experiments were carried out at the Argonne Tandem Linac Accelerator System (ATLAS) in inverse kinematics. Both of these reactions are relevant for nuclear astrophysics and were measured at low energies. The beam energies for the $^{17}$F and $^{17}$O beams were about 50 MeV and 35 MeV, respectively. The $^{17}$F beam is radioactive and was produced by using the in-flight technique via the $^{16}$O$(d,n)^{17}$F reaction. The maximum beam intensity of the $^{17}$F beam was about 3000 pps. The MUSIC detector was filled with  pure $^4$He gas at room temperature and a pressure of 400 Torr for the $^{17}$F$+\alpha$ case and 206 Torr  for the $^{17}$O$+\alpha$ case.  In both cases, the data acquisition system was based on analog electronics. The physics motivation, final cross section, and more details of the $^{17}$F$(\alpha,p)^{20}$Ne reaction will be presented in a future publication. The cross section of the $^{17}$O$(\alpha,n)^{20}$Ne reaction and more details of the experiment have been published in \cite{AvilaNIM16}.

\section{Simulated data}
\label{s:simulation}

The computer program \musicsim~\cite{musicsim} was  written to simulate the detector response based on the kinematics of the reactions occurring in the detector, the geometry of the detector, and the energy loss of the particles as they travel through the gas (detector medium). The simulation takes as inputs the kinetic energy of the beam particles as they enter the gas volume, the beam energy spread, the random noise for all the detector segments, the ion species---e.g., $^{12}$C for a Carbon-12 nucleus---for initial (beam and target), and final (i.e., after a nuclear reaction) particles. To simulate a nuclear reaction in a specific detector strip, the program randomly selects a point of interaction along the trajectory of the beam particle (within the selected strip) and calculates the energy loss of the beam as it travels from the entrance window to the interaction point. To calculate the energy loss, one can use stopping powers from different tables, such as the Ziegler \cite{SRIM} and ATIMA \cite{ATIMA}. After the energy is calculated, the simulation program randomly selects an angle in the center of mass and calculates the kinematics of the outgoing particles, then boosts the 4-momenta of the outgoing particles from the center of mass to the laboratory frame through a Lorentz transformation. Using the angles and energies of the different particles (in the laboratory frame), the simulation program uses energy loss tables to calculate the energy deposited per strip, as it is measured in the MUSIC detector.  

\begin{figure}
    \centering
    \begin{tabular}{c}
        \includegraphics[scale=0.35]{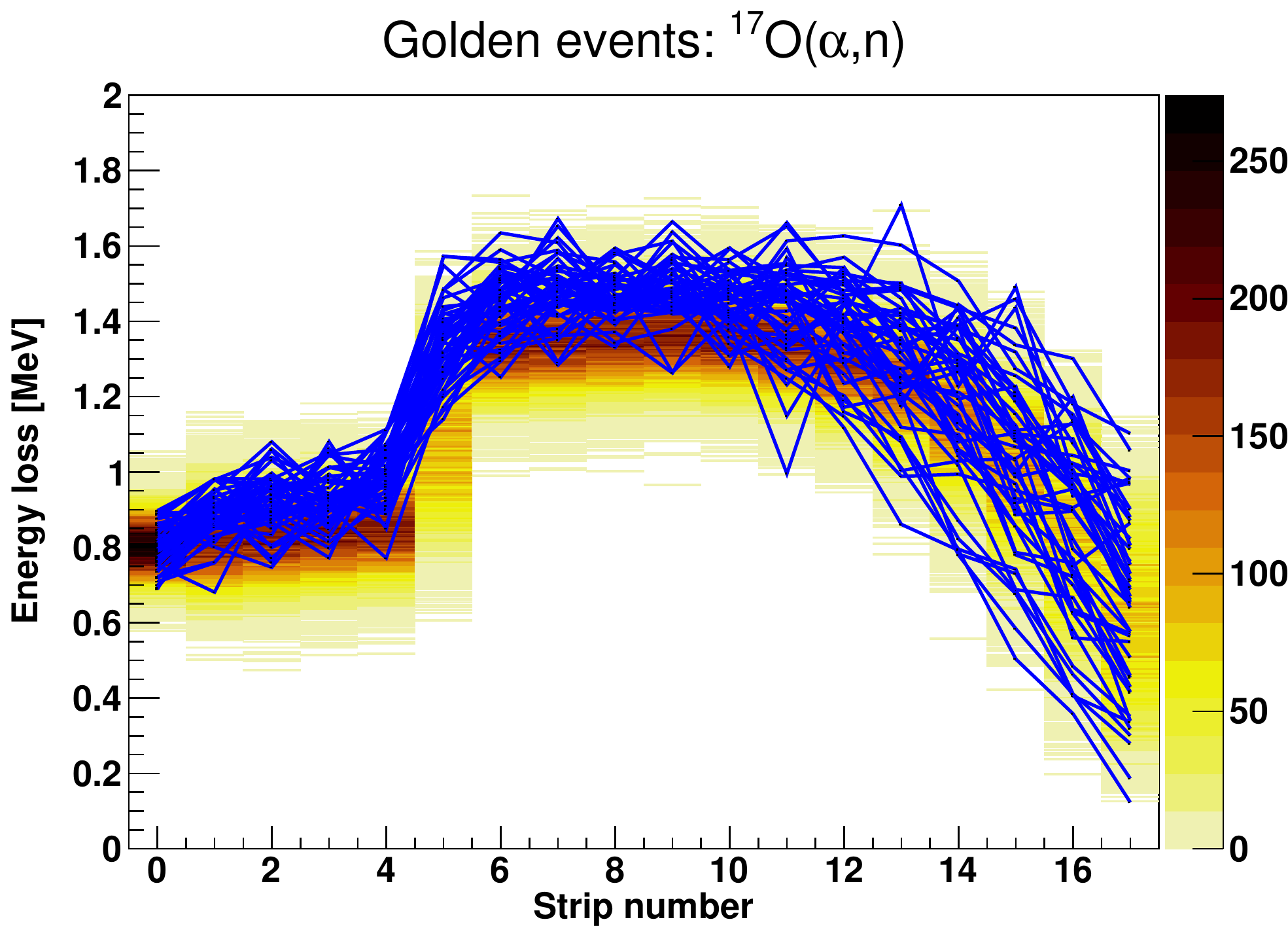}\\
        \includegraphics[scale=0.35]{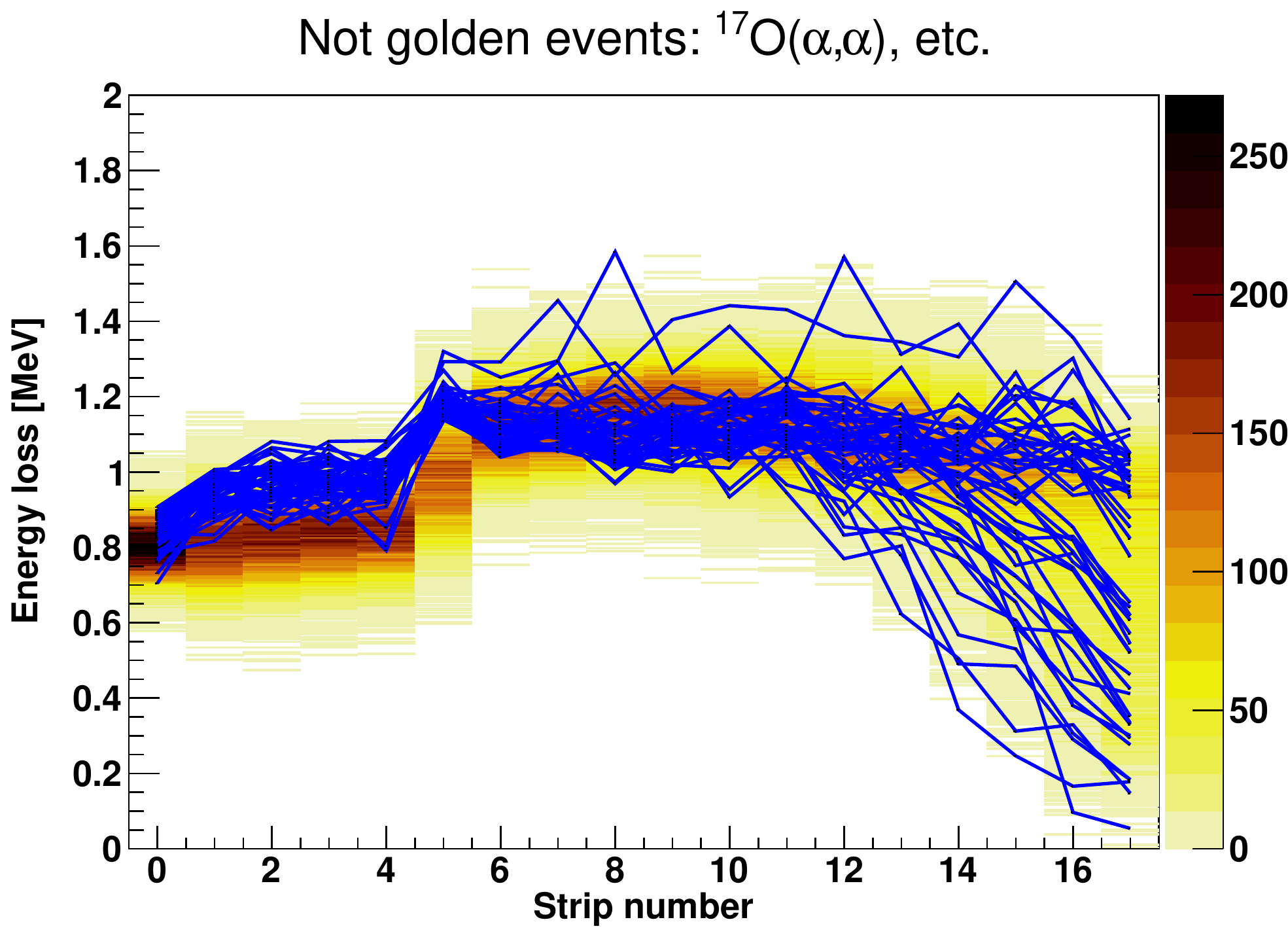} \\
        \includegraphics[scale=0.35]{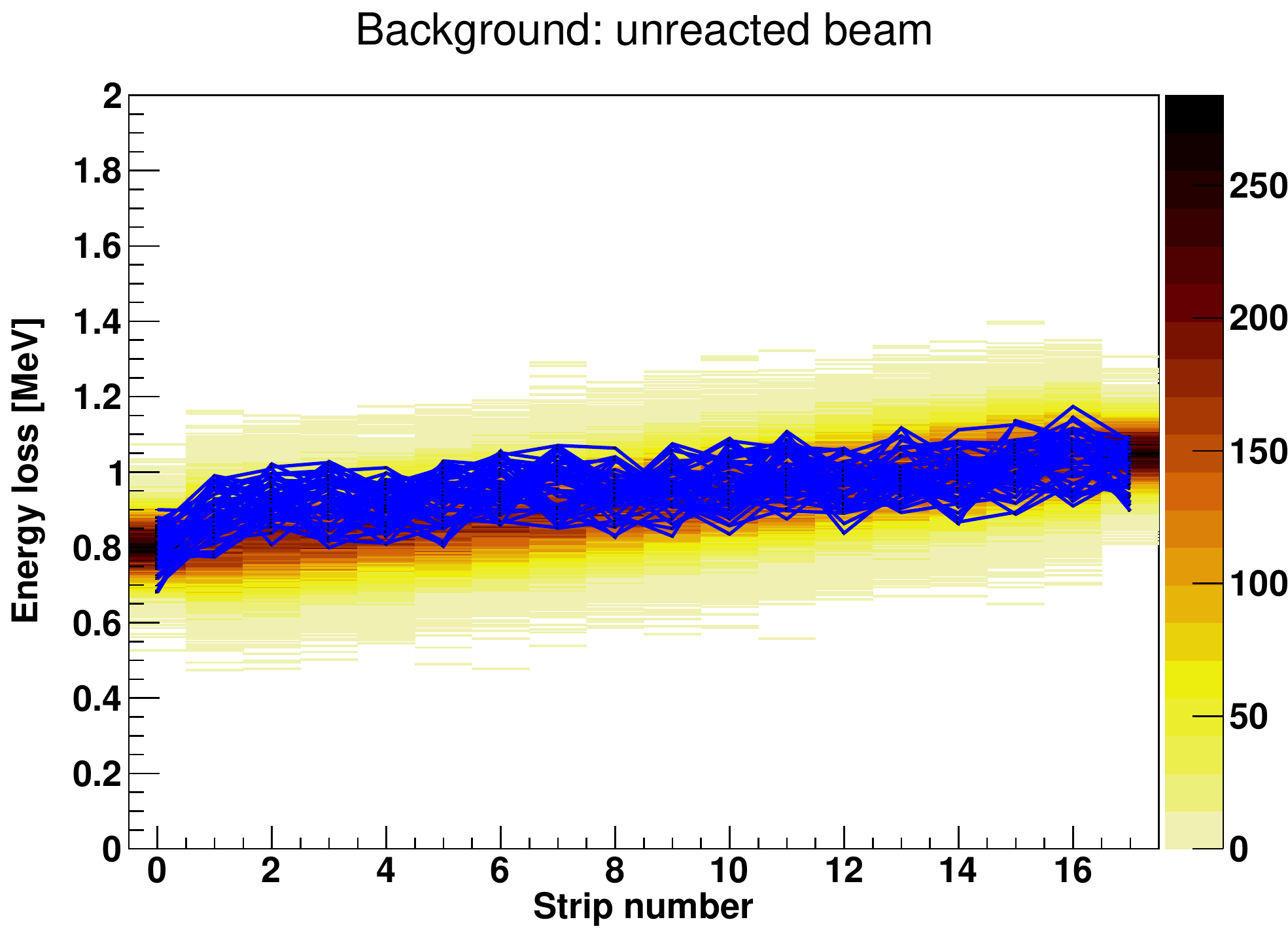}
        
    \end{tabular}
    \caption{\label{fig:simVal} Comparison of simulated (histograms) and experimental (blue lines) data for the $^{17}$O$+\alpha$ case. Several individual experimental traces are shown with blue lines and are displayed over the histograms for (bottom) unreacted $^{17}$O beam and for events occurring in strip 5 classified as ``not golden events'' (middle), and ``golden events'' (top).}
\end{figure}

Figure \ref{fig:simVal}  shows examples of simulated (histogram on the background) and experimental (blue lines on the foreground) data for the $^{17}$O$+\alpha$ case. In this  data visualization, the total energy deposited in each strip is plotted as a function of the strip number. For the strips with left/right segmentation, the y-axis is composed of the sum of the energy deposited in the left and right segments. Three cases are shown in the figure: golden events (top), which in this case correspond to events from $^{17}$O$(\alpha,n)^{20}$Ne reactions; not golden events, in this case $^{17}$O$(\alpha,\alpha)^{17}$O reactions; and background (bottom), for example from the unreacted $^{17}$O beam, which are the most abundant background events but experimentally could include other spurious events and noise. For the  histograms (from simulation) the bins with dark color indicate a higher frequency than those with lighter colors. For the $^{17}$O$(\alpha,n)$ and $^{17}$O$(\alpha,\alpha)$ simulations, an isotropic angular distribution (in the center of mass) has been assumed; however, the actual physical distribution is unlikely to be isotropic.

Empirically, separation between golden events (GE) and not golden events (not-GE) becomes increasingly more challenging as we probe strips located deeper into the detector, where the energy loss differences are smaller. The separation between GE and not-GE also is expected to diminish when the outgoing heavy ion differs by only one proton from the beam, as in the case of $^{17}$F$(\alpha,p)^{20}$Ne.

\section{AI/ML methods for MUSIC}
\label{s:methods}
The experimental data obtained with the MUSIC detector can be denoted as a matrix $\mathcal{X} \in \mathbb{R}^{n \times p}$ that is a collection of events from the detector, where $n$ is the total number of events data from the detector and $p$ is the dimension of the each event vector containing the particle's energy loss in each detector strip.
As indicated in Sec.~\ref{s:music}, the nominal value of $p$ is $18$, and the typical value of $n$ is $\sim10^{9}$.  We divide $\mathcal{X}$ into background ($\mathcal{X}_{\mathrm{back}}),$ golden events (events of interest, $\mathcal{X}_{GE}$), and other events ($\mathcal{X}_{OE}$ with $\mathcal{X}=\mathcal{X}_{\mathrm{back}} \cup \mathcal{X}_{GE} \cup \mathcal{X}_{OE}$). The exact proportion of the background, golden, and other events is unknown. The goal of the analysis is to extract $\mathcal{X}_{GE}$ from $\mathcal{X}$ under the assumption that the total number of background events is significantly larger than for golden events. For example, in the case of $\alpha$-induced reactions on $^{17}$F, the golden events $\mathcal{X}_{GE}$ are $(\alpha,p)$ events. Similarly, $(\alpha,n)$ events are the golden events for $\alpha$-induced reactions on $^{17}$O.  

The simulation data is denoted as $\mathcal{S}$, which can be divided into three groups: $\mathcal{S}_{\mathrm{back}}$, typical background events (mostly from unreacted beam); $\mathcal{S}_{GE}$, golden events~(events of interest); and $\mathcal{S}_{OE}$, the other events. These samples are obtained with the simulation code described in Sec.~\ref{s:simulation}. 

\begin{figure}
    \centering
    \includegraphics[width = \textwidth]{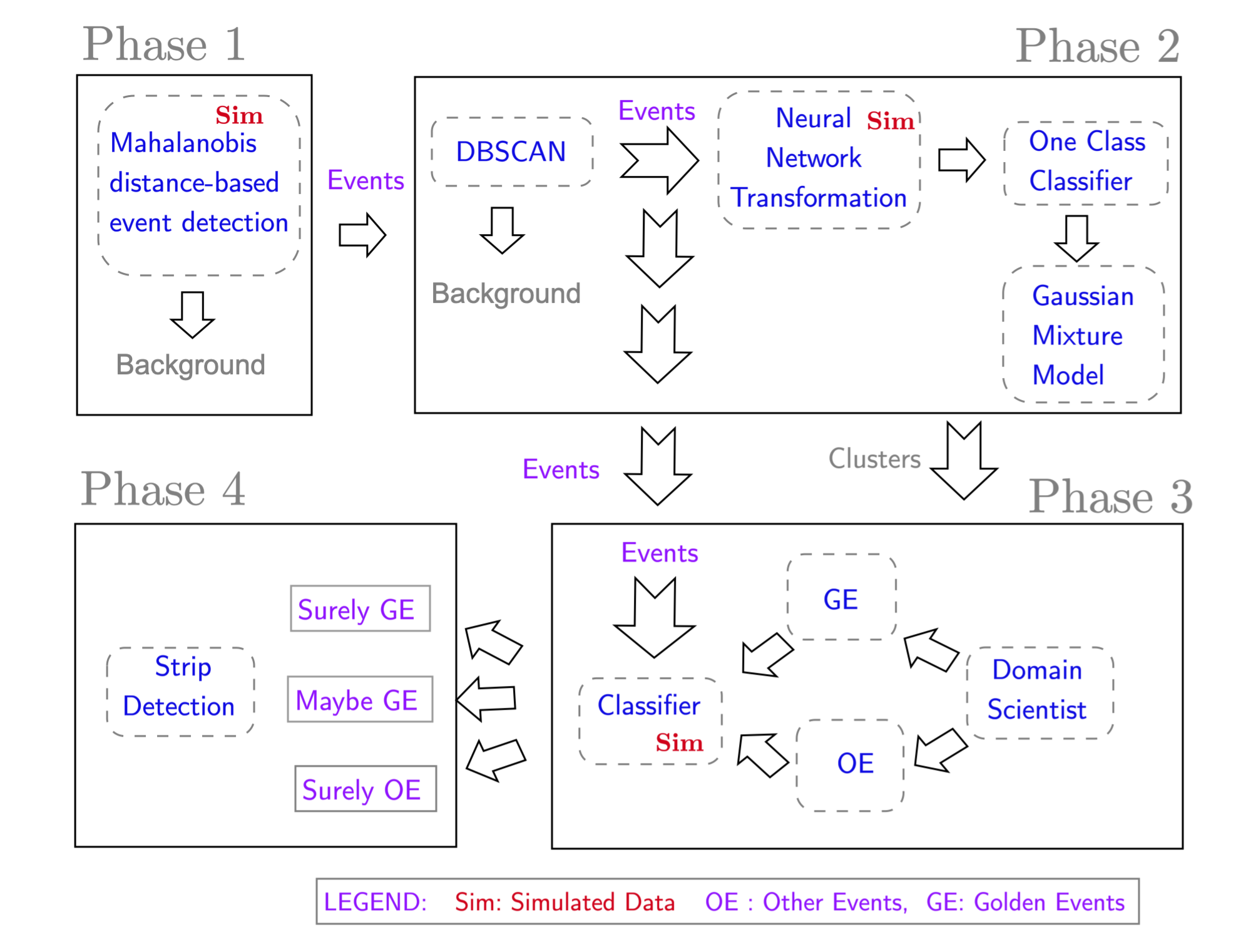}
    \caption{Schematic of our four-phase approach to golden event detection in MUSIC detector.}
    \label{fig:process}
\end{figure}

The methodology to extract $\mathcal{X}_{GE}$ comprises four phases. See Fig.~\ref{fig:process} for a high-level overview. 
\begin{description}
    \item[Phase 1:] Remove easy-to-identify background events from $\mathcal{X}$ using statistical techniques.
    \item[Phase 2:] Use a neural network transformation to remove background events that are similar to golden events.
    \item[Phase 3:]  Design a classifier using input from domain scientists to extract three clusters that indicate events that are surely golden~(``Surely GE"); are surely other events~(``Surely OE"); and may or may not be golden events~(``Maybe GE").
    \item[Phase 4:] Using the rate of change of energies at each strip, determine the strips corresponding to the location~(detector segment, signified by the strip number) where the event took place.
\end{description} 

Below, each phase is discussed in detail.
\subsection{Phase 1: Removing background events}
The total number of events in $\mathcal{X}$ is around $10^9$. The majority of these  (roughly  $10^5$) are background events with traces similar to those shown in the bottom panel of  Fig.~\ref{fig:simVal}. The easy-to-identify background events are removed by using a distance -- a measure-based detection. 

The Mahalanobis distance (MD) is a  measure of the distance between a point and a distribution~\cite{soylemezoglu2011mahalanobis}.
Typically, one sample is denoted as reference~($\mathcal{R}$ with mean $\bar{r}$ and covariance matrix $\Sigma$) and one sample as test~(in this case $\mathcal{X}$). MD then quantifies the difference between reference sample $\mathcal{R}$ and a point~$x \in \mathcal{X}$ such that the MD, $d(x,\mathcal{R})$, is given as
	\begin{equation} \label{1} 
	\begin{aligned} 
	d(x,\mathcal{R})=\|\Sigma^{-1/2}(x-\bar{r})\|_{2}^{2},
	\end{aligned} 
	\end{equation} 
where $\Sigma^{-1/2}$ denotes the Cholesky decomposition \cite{johnson2014applied} of the covariance  matrix and $\|.\|_{2}^{2}$ is the squared 2-norm. In practice, $\Sigma $ is replaced by its sample estimate $\hat{\Sigma }$. Typically, for any normally distributed reference and a test point $x$, $d(x,\mathcal{R})<1$ would imply that $x$ belongs to the same distribution as the reference. Therefore, with the goal of removing background events, one may obtain a reference sample that is representative of background events and designate any event $x \in \mathcal{X}$ as background, if $d(x,\mathcal{R})$ is less than a threshold $d_B.$  

Formally, the background events can be removed by testing the null hypothesis $$H_0: x \sim p(\mathcal{R}),$$  where $\mathcal{R}$ is the reference sample and $p(\mathcal{R})$ is a distribution over it.  In order to test $H_0$ based on MD, a distance $d(x,\mathcal{R})$ is calculated for each $x \in \mathcal{X}.$ 
Therefore, for rows $x_i$ in $\mathcal{X}$, where $i=1, \cdots, n$, the distances are denoted as $\mathcal{D} =\{d_1, d_2, \cdots d_{N}\}$. With sample $\mathcal{D},$ a sampling distribution $p(\mathcal{D})$ and a distance threshold $(d_B)$, $\forall d \in \mathcal{D},$ the null and alternate hypothesis can be described as 
\begin{align}
    H_0: d \leq d_B \quad \quad  H_1: d > d_B .
\end{align} 
Satisfying the null hypothesis indicates that $x \sim p(\mathcal{R})$~(i.e., $x$ is a background event), and satisfying the alternate hypothesis indicates the $x$ may be a golden event. To perform this hypothesis testing, one must obtain an accurate $\mathcal{R}$ reference sample and an accurate $d_B$ distance threshold.

\textit{Reference Sample:} With the assumption that the total number of background events in $\mathcal{X}$ far exceeds the total number of golden events, a uniform sample from $\mathcal{X}$ denoted as $\tilde{X}_{b}$ provides a reasonable representation of the background events. Furthermore, a simulation sample~$\mathcal{S}_{\mathrm{back}}$ corresponding to the background events is also provided, to be used along with $\tilde{X}_{b}$ to constitute the reference sample---$\mathcal{R} = \mathcal{S}_{\mathrm{back}} \cup \tilde{X}_{b}$.

\textit{Distance Threshold:} The distance threshold is obtained by designating $d_B = P_r$, where $P_r$ denotes the $r$th percentile of $p(\mathcal{D}).$ A  common choice of $r$ is $99$ such that $P_r$ denotes the $99$th percentile. We note that a low value of $r$ yields a large type one error~(i.e., more background events are misclassified as potential golden events). All the $x \in \mathcal{X}$ that satisfies the null hypothesis are designated as background events;  the rest of the events are collectively denoted as $\mathcal{X}_{PE}$.

\subsection{Phase 2- Outlier detection}
As previously mentioned, a large number of events in $\mathcal{X}$ are background events originating from unreacted beam particles.  Since the reference sample is obtained by uniformly sampling $\mathcal{X},$  it is a good representation of an unreacted beam.  MD-based  detection is limited to removing such background events. Thus, $\mathcal{X}_{PE}$  comprises golden events, other events and many background events that are similar in their trace to golden events or other events. Therefore, to precisely extract golden events, one must remove other events and leftover background events from $\mathcal{X}_{PE}$. We will first utilize DBSCAN~(density-based spatial clustering of applications with noise) to remove other events and develop a neural-network-based transformation to reduce similarities between golden events and background events.

\subsubsection{DBSCAN -- Density-Based Spatial Clustering of Applications with Noise} \label{sec: dbs}
DBSCAN~\cite{khan2014dbscan} is a nonparametric algorithm where, given a set of points, DBSCAN finds points that are closely packed~(points with many nearby neighbors---the high-density region) and points whose nearest neighbors are too far away~(the low-density region). Under the assumption that all the ``OE'' will cluster together in the high-density regions, DBSCAN is used to extract points that are closely packed, and these points are designated as $\mathcal{X}_{OE}.$ Similarly, the events that lie in the low-density region are designated as $\mathcal{X}_{PGE}$, which is referred as the set of all potential golden events such that $\mathcal{X}_{PGE}$ is a collection of golden events and background events that are similar in their trace to golden events. In particular, we would like to design a classifier that can take as input $\mathcal{X}_{PGE}$ and label them as other events and golden events. To design this classifier, however, we need to obtain labeled data. The next two steps are therefore employed to extract that core subset of data that can be labeled by the domain scientists.

\subsubsection{Neural Network-Based Transformation}\label{sec:NNt}
It has been empirically observed that in cases such as the one for $^{17}$F$(\alpha,p)$ reactions,  several background events  are  similar to golden events. In such scenarios one can perform a transformation that  reduces the similarity between golden and hard-to-identify background events. In order to construct this transformation,  an arbitrary neural network~(NN) can be employed. NNs comprise layers where each layer is built by using a linear transformation employing weights and a nonlinear function. Typically, the weights of the NN can be trained by minimizing a loss function. The weights are collectively denoted as $\theta$ such that the NN can be denoted as $f_{\theta}.$ In this work, the network is trained to expand the differences between golden and background events with the goal of minimizing a cost $J(\theta)$ such that $$\theta^{*} = \mathrm{argmin}_{\theta} J(\theta).$$ 

The cost $J(\theta)$ is designed to quantify the differences between golden and background events. To this end, three sample sets are defined: the background~$\mathcal{T}_{\mathrm{back}} = \mathcal{S}_{\mathrm{back}} \cup \mathcal{R}$; golden events~$\mathcal{T}_{GE} = \mathcal{S}_{GE}$; and other events ~$\mathcal{T}_{OE} = \mathcal{S}_{OE} \cup \mathcal{R}_{OE},$ where $\mathcal{R}_{OE}$ is a sample from $\mathcal{X}_{OE}$ obtained at the end of Subsection \ref{sec: dbs}.  Equipped with these samples, we define $J(\theta)$  as follows.
\begin{align}
    J(\theta) = \mathbb{E}_{x \in \mathcal{T}_{\mathrm{back}}, y \in \mathcal{T}_{GE}, z \in \mathcal{T}_{OE}} \bigg[ \|z-f_{\theta}(x)\|_{2}^{2} - \gamma \big[ \| x-f_{\theta}(y) \|_{2}^{2} +\|z-f_{\theta}(y) \|_{2}^{2} \big] \bigg]
    \label{eq: object}
\end{align} 
where $\mathbb{E}$ is the expected value operator. In this cost function, the distance between a golden event and reference/other events is increased by maximizing the term $[ \| x-f_{\theta}(y) \|_{2}^{2} +\|z-f_{\theta}(y) \|_{2}^{2}]$, and the distance between the other-events and background sample is reduced by minimizing the term $\|z-f_{\theta}(x)\|_{2}^{2}$. The $\theta^*$ that satisfies the objective Eq.~\eqref{eq: object} increases the distance between golden events and the rest (other events and background). This transformation can be estimated by using standard learning techniques such as stochastic gradient descent~\cite{hardt2016train}. Once this transformation is learned, $f({\theta^*})$ is used on all $\mathcal{X}_{PGE}$ such that $\mathcal{X}^T_{PGE}= f_{\theta^*}(\mathcal{X}_{PGE}).$  From here on, the application of the NN transformation is denoted by a superscript $^T$. Once we transform all $\mathcal{X}_{PGE}$, a nearest-neighbor-based scheme is utilized to remove most of the leftover background events and other events. 

\subsubsection{Nearest-Neighbor-Based Outlier Detection}
The events from $\mathcal{X}^T_{PGE}$ that look similar to background and other events are removed by using a nearest-neighbor-driven search. In order to efficiently perform this search, a tree-based approach~\cite{ram2019revisiting} is employed, and a k-d tree~(hereafter denoted as $Kdt$) is defined on $\mathcal{R}^T,$ where $\mathcal{R}^T$ is the reference sample transformed using the NN (refer to Section \ref{sec:NNt}). With this tree, a nearest-neighbor search algorithm~\cite{bishop2006pattern} is executed to obtain k-neighbors from $\mathcal{R}^T$ that are closest to a given $x \in \mathcal{X}^T_{PGE}.$ Based on these k neighbors, one determines whether an event is golden or not according to the average distance between $x$ and different neighbors. 

Formally, for every  $x \in \mathcal{X}^T_{PGE},$ $k$ neighbors are first determined by using~$Kdt.$ Next, the average distance between~$x$ and the $k$ neighbors~($d_{\mathrm{avg}}$) is evaluated, where $d_{\mathrm{avg}} = (\sum_{i=1}^{k} d_i)/k)$ with $ d_i$ being the distance between each neighbor and $x.$  Based on a threshold $d_{\mathrm{thre}},$ the following hypothesis is tested for each $x$.
\begin{align}
    H_0: d \leq d_{\mathrm{thre}} \implies x \sim p(\mathcal{R}^{T}) \quad 
    H_1: d > d_{\mathrm{thre}} \implies x \sim p(\mathcal{T}_{GE})
\end{align}
For any $x,$ if the null hypothesis is accepted, a background event is detected;  if the alternate hypothesis is accepted,  a golden event is detected. The strategy utilized in phase 1 can be used to determine $d_{\mathrm{thre}}.$ Based on the results of this hypothesis testing, outliers are extracted and designated as $O.$ The set of outliers~$(O)$ are then clustered into multiple groups by using the Gaussian mixture model and given to the domain scientists for labeling. The clusters obtained at the end of last phase are labeled by the domain scientists as either an other events cluster~(a collection of other events clusters is denoted as $\tilde{\mathcal{X}}_{OE}$) or a golden events cluster~(a collection of golden events cluster is denoted as $\tilde{\mathcal{X}}_{GE}$). These two class samples can now be employed to classify $\mathcal{X}_{PGE}$ and obtain the final clusters.

\subsection{Phase 3: Classification}
With these two class samples, $\tilde{\mathcal{X}}_{GE}$ and $\tilde{\mathcal{X}}_{OE}$, a  two-step classification method is designed to determine the three event groups ``Surely GE," ``Maybe GE," and ``Surely OE." Almost any classification method~\cite{bishop2006pattern} could be used to determine the three event groups since the  training data sample corresponding to GE and Not GE is now available. In our preliminary analysis, the results  recorded in Table~\ref{tab:out} in Sec.~\ref{s:results} show that the most  GE's were detected by One-class SVM~\cite{scholkopf1999support}  albeit with many false positives. Since most of the background events~(the source of false positives) have been removed and a good training sample is now available, one-class SVM is an ideal choice as a classifier in the two-step classification procedure.

In the first step, the ``Surely GE" class is extracted by using an SVM classifier~\cite{bishop2006pattern} that is designed to separate ``Surely GE"~$(\mathcal{L}_{GE})$ from $\mathcal{X}_{PGE}$. Specifically, the GE class for the classifier is defined as $\tilde{\mathcal{X}}_{GE} \cup \mathcal{S}_{GE}$ and the OE class as $\tilde{\mathcal{X}}_{OE} \cup \mathcal{S}_{OE} \cup \mathcal{S}_{\mathrm{back}}.$ 

Second, the ``Maybe GE" class is extracted from $\mathcal{X}_{PGE} | \mathcal{L}_{GE} $, that is, ~$\mathcal{X}_{PGE}$ with $\mathcal{L}_{GE}$ removed.  The training data for the GE class is redefined as  $\mathcal{S}_{GE} \cup \mathcal{L}_{GE}, $ and the OE class is provided as $\tilde{\mathcal{X}}_{OE} \cup \mathcal{S}_{OE} \cup \mathcal{S}_{\mathrm{back}}.$  Using this new training data, a classifier is used to extract all the remaining golden events from  $\mathcal{X}_{PGE}$, where  $\mathcal{L}_{GE}$ are removed.  The result of this classifier is designated as ``Maybe GE," and all the remaining events  are designated as the ``Surely OE" class. 

\subsection{Phase 4: Strip Detection}
The final step is to determine at which detector segment a golden event occurred.
Experimentally, when a nuclear reaction occurs within one MUSIC strip, the measured energy deposited in that strip, and the following strips change in a way that depends on the atomic number of the resulting ions and on their kinematics. In the $(\alpha,p)$ and $(\alpha,n)$ cases, an event is said to happen at strip $i$ if there occurs a significantly large increment in the energy deposited in the strip.
This characteristic ``jump"  can be detected by using the gradient of the energy level with respect to the strip number~(that is, the gradient is maximum at the strip corresponding to the characteristic jump).  Therefore, for any data point $x,$ the gradient at each strip for $x = \{ x_i, i = 1 \cdots p \}$ is given as $G^x_i = \frac{\partial x}{\partial i } \approx |x_{i} - x_{i-1}|.$  The strip number $s_x$ for $x$ then is given as $s_x = \mathrm{argmax}\; G_x$ such that $G_x = \{ G^x_i, i = 1, \cdots p \}.$ Equipped with this algorithm, we determine the strip numbers for each event categorized at the end of the last phase, in other words, an event in the``Surely GE" class, ``Maybe GE" class, and ``Surely OE" class.

\section{Results and discussion}
\label{s:results}

The new method described above  was developed for the strip-wise classification of $(\alpha,p)$ and $(\alpha,n)$ reactions taking place in the MUSIC detector. We have applied the  method  to two experimental data sets:  the $^{17}$F$+\alpha$ data set, where $^{17}$F$(\alpha,p)^{20}$Ne reactions are the ``golden events,'' and the $^{17}$O$+\alpha$ data set for the classification of $^{17}$O$(\alpha,n)^{20}$Ne events. In the former case,  no published experimental data exists; nevertheless, a preliminary analysis using traditional classification methods to tag the $^{17}$F$+\alpha$ events has been performed so that the performance of the new method can be assessed. For the latter case, the results are compared with the experimental data published in \cite{AvilaNIM16}. Based on the physics of the detector response (see Sec.~\ref{s:music}), typically the principal differences in the MUSIC traces are enhanced when more protons are removed from or added to the beam particles ($\Delta Z$). Therefore, it is expected that for the $^{17}$F$(\alpha,p)^{20}$Ne case ($\Delta Z=1$) the classification of GE would be more challenging than for the $^{17}$O$(\alpha,n)^{20}$Ne case ($\Delta Z=2$).
In this section we discuss the results obtained by applying the new method to these two MUSIC data sets.

\subsection{Classification of $^{17}$F$(\alpha,p)^{20}$Ne events}

\begin{figure}
    \centering
    \begin{tabular}{c}
        \includegraphics[scale=0.5]{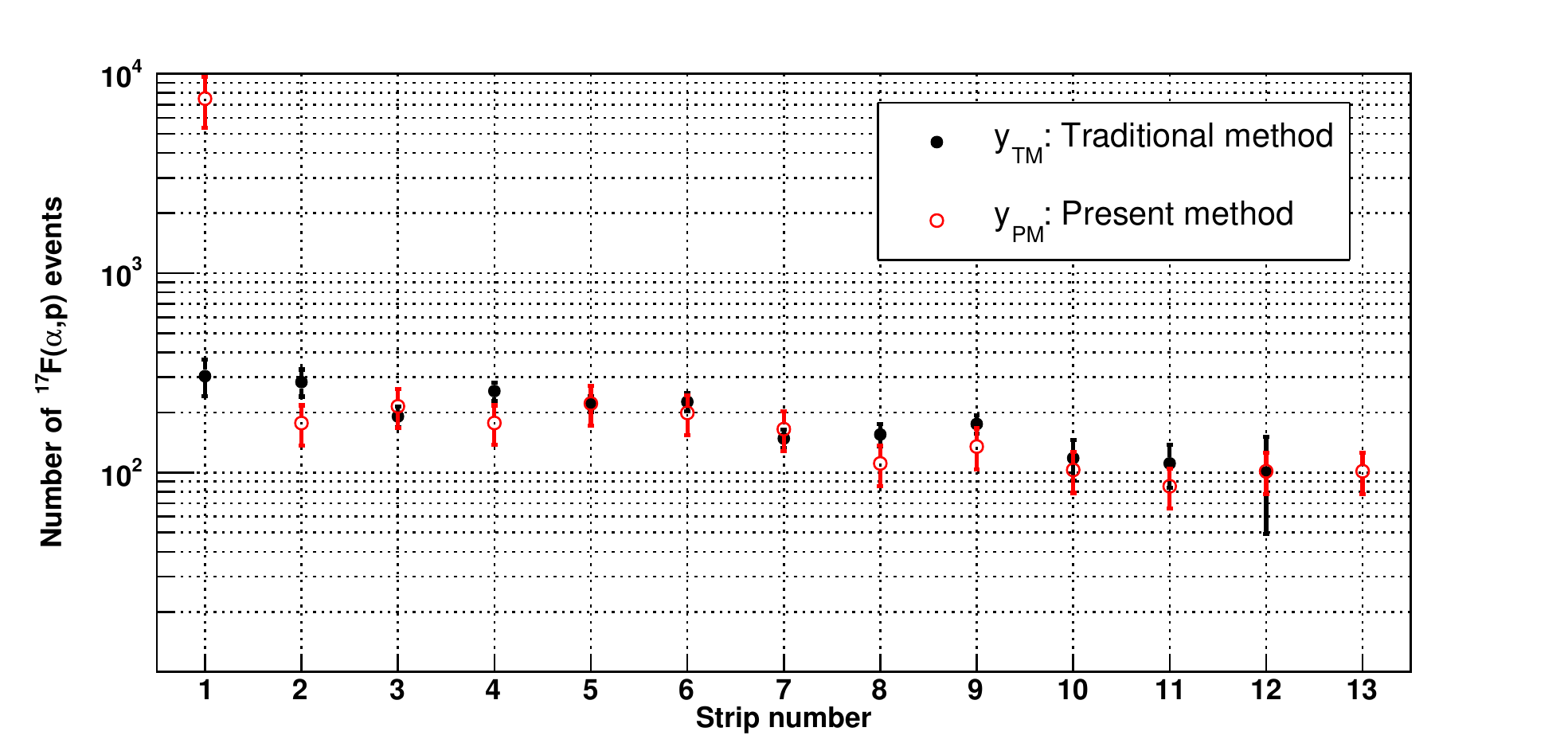} \\
        \includegraphics[scale=0.5]{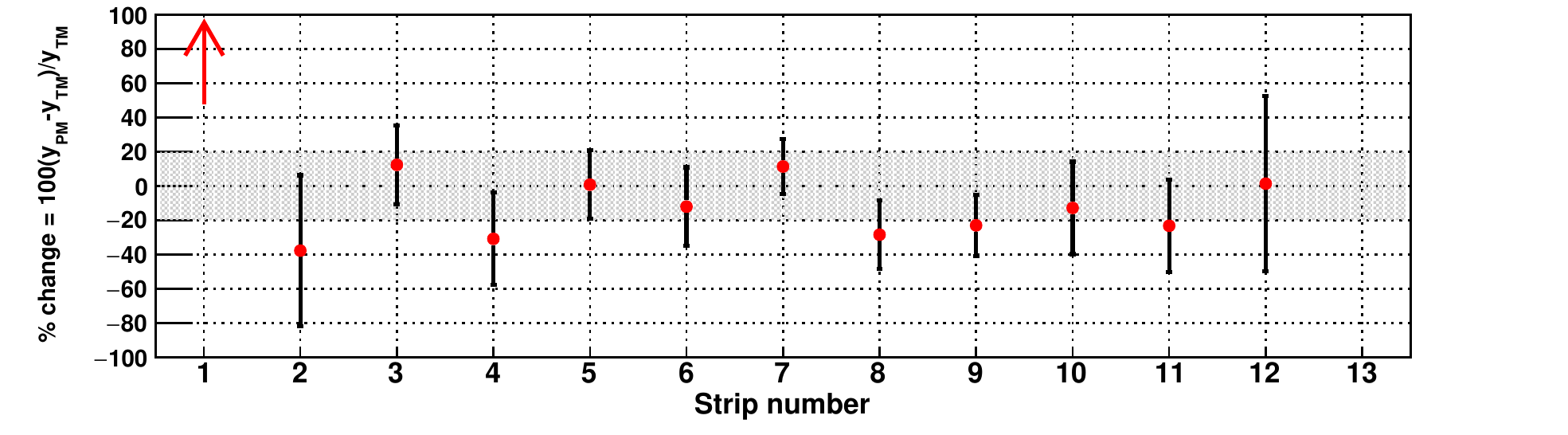}
    \end{tabular}
    
    \caption{\label{fig:compareF} Comparison of results for the classification of $^{17}$F$(\alpha,p)$ events with the traditional technique (black points) or with the present method (red open circles). See text.
    }
\end{figure}

In the top panel of Fig.~\ref{fig:compareF}, we show the number of events classified as ``Surely GE,'' in this case $^{17}$F$(\alpha,p)$ events, plotted as a function of the MUSIC strip number. Such event classification is a key step in the extraction of the $^{17}$F$(\alpha,p)$ reaction cross section, which is the quantity of interest for the domain scientists. In Fig.~\ref{fig:compareF}, the black points are obtained by using the standard technique (traditional method), and the red open circles are obtained with the new (present) method. In the top panel the error bars of black points include statistical and systematic uncertainties while the error bars for the red  points relate to the aleatoric and epistemic uncertainties of the new method (see Sec.~\ref{s:uncert}). In the bottom panel we show the percent change in the number of events classified as ``Surely GE'' using the new method with respect to the number of $^{17}$F$(\alpha,p)$ events classified using the traditional method as a function of the strip number. The shaded area encompasses the region of $\pm 20\%$. Error bars in the bottom graphs are the experimental ones. The up arrows in strip 1 indicate that the corresponding \% change values are off the scale.

The largest disagreement is found in the classification of GE for strip 1. Such a dramatic increment in the number of counts, going from strip 2 to strip 1, is not expected from the experimental point of view, and it is likely not physical. This suggests that, for the new method, refinements are needed to distinguish strip-1 GE from not-GE or from background signals. Intuitively, this is not too surprising since there is not a robust baseline---in this case the baseline is just formed by the strip 0 data--upon which to signature ``jump'' in the MUSIC traces would be apparent. 
Empirically, we also find challenging the classification of GE in the strips near the edges of the detector (in this case, strips 1 and 2, and beyond strip 10) when using traditional methods. Therefore, larger systematic uncertainties are expected near the edges of the detector.

\begin{table}[!tbh]
    \centering
    \begin{tabular}{c|cc}
    $^{17}$F$(\alpha,p)$  & GE pres.~meth.&  Not-GE pres.~meth. \\ \hline
    GE trad.~meth.     & 1825(32)         &  186(32)     \\
    Not-GE trad.~meth. & 7968(1311)        &  569882(1311)
    \end{tabular}
    \caption{Confusion matrix for the $^{17}$F$(\alpha,p)$ case showing the mean number of events with standard deviation in parentheses. 
    }
    \label{tab:17_F}
\end{table}

Table \ref{tab:17_F} shows the confusion matrix for two types of events classified by the new method  and traditional methods. Again, in this case the golden events (GE) correspond to  $^{17}$F$(\alpha,p)$ reactions, and not-GE are events that are not considered background but rather other types of reactions such as $(\alpha,\alpha)$ events (background is mostly unreacted beam events). Physically, the not-GE events likely contain some $\alpha$-scattering events but might also contain spurious events originating from electronics noise in some of the detector channels.  In the table we show the mean number of events with standard deviation in parentheses. In order to evaluate these quantities, the four phases of the new method were executed for 50 repetitions. The values are rounded off to the nearest integer. 

As mentioned above, the large number of false positives, namely, the 7968 events in Table~\ref{tab:17_F}, originate from the disagreement in classification of events in strip 1. When strip 1 events are removed from the analysis, the number of false positives is dramatically reduced, from 7968 to 279 events, while the number of true positives remains high, changing from 1825 to 1736 events. The number of false negatives, namely, the 186 events in Table~\ref{tab:17_F}, is less than 10\% of the total number of GE classified using traditional methods. We compare the accuracy of the present method with other outlier detection methods in Sec.~\ref{s:otherMLmeth}.

\subsection{Classification of $^{17}$O$(\alpha,n)^{20}$Ne events}

\begin{figure}[h]
    \centering
    \begin{tabular}{c}
        \includegraphics[scale=0.5]{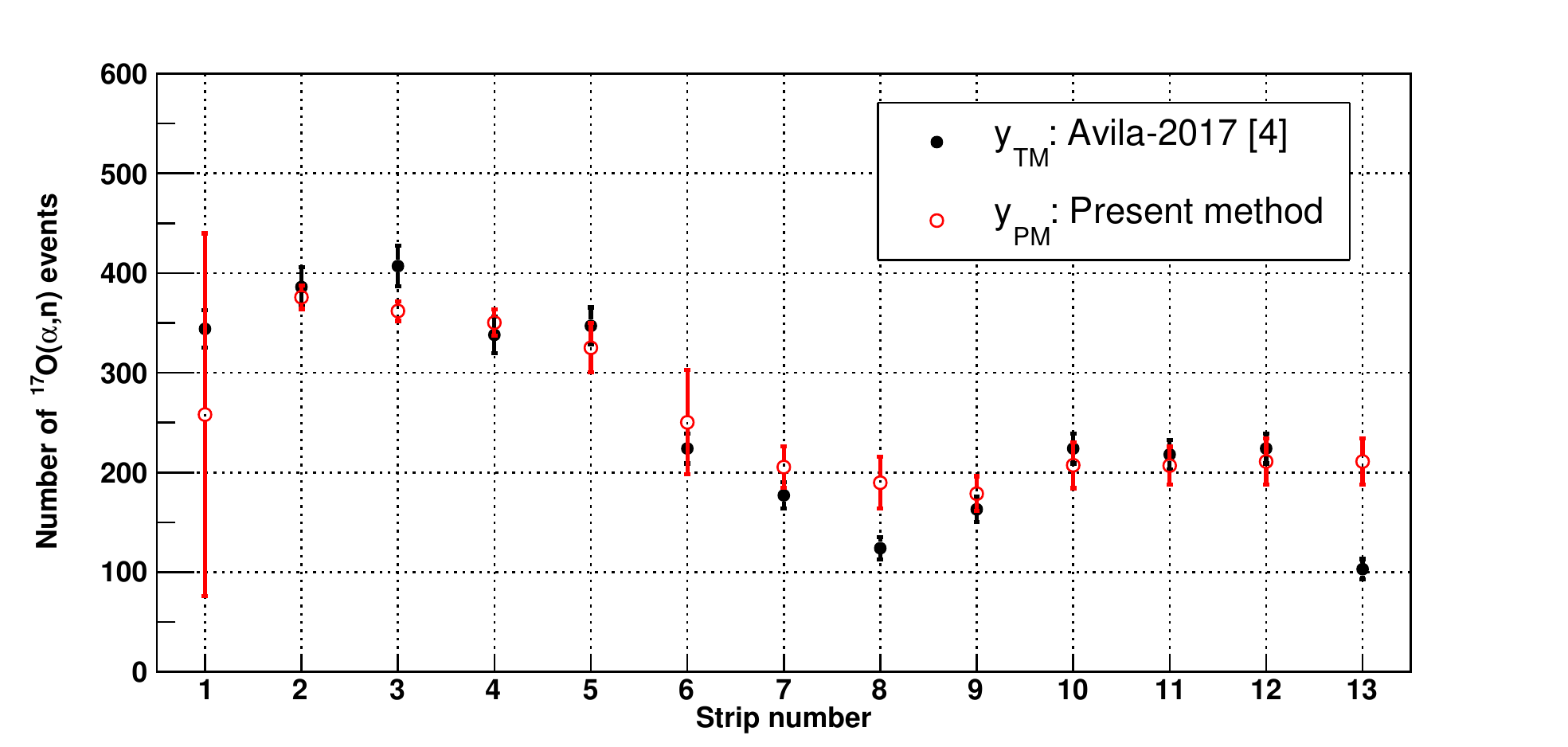} \\
        \includegraphics[scale=0.5]{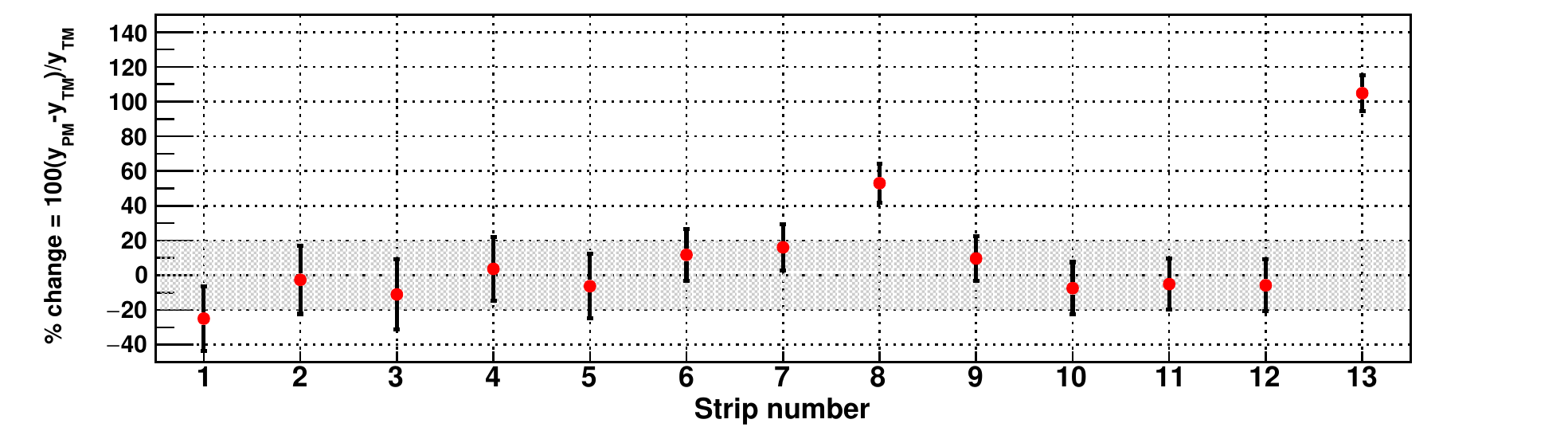}
    \end{tabular}
    \caption{\label{fig:compare} Similar to Fig.~\ref{fig:compareF} but for a different data set, where the ``golden events'' correspond to  $^{17}$O$(\alpha,n)$ reactions. The experimental data is from \cite{AvilaNIM16}.}
\end{figure}

In Fig.~\ref{fig:compare} we compare the results for the classification of $^{17}$O$(\alpha,n)$ events obtained from the traditional method and the new method. As previously mentioned,  the experimental data was published in~\cite{AvilaNIM16}. In the top panel, we show the number of events classified as $^{17}$O$(\alpha,n)$ by the standard technique (black circles) or by the present method (red open circles) plotted as a function of the MUSIC strip number. 
As in the preceding section, the error bars of the black points are only statistical, while the error bars for the red points relate to the aleatoric and epistemic uncertainties of the present model (see Sec.~\ref{s:uncert}). In the bottom panel, we show the percent change of number of events classified as ``Surely GE,'' using the new method, with respect to the number of $^{17}$O$(\alpha,n)$ events classified using traditional technique as a function of the strip number. The shaded area encompasses the region of $\pm 20\%$.

In this case the largest disagreement is found in the classification of GE for strips 8 and 13. As mentioned before, the classification of GE in the strips near the edges of the detector is more challenging than away from the edges. Therefore, the disagreement in strip 13 is not surprising. On the other hand, we currently do not fully understand the disagreement in strip 8. However, the cross section for that strip reported in \cite{AvilaNIM16} is smaller than the one obtained in an earlier experimental work \cite{Bai73}, as seen in Fig.~5 of Ref.~\cite{AvilaNIM16}. A separate investigation to see whether the prior data analysis misclassified some events in this region could help us understand this difference, but such effort is beyond the scope of this paper.


Table \ref{tab:17_O} shows the confusion matrix for GE and not-GE classified by the new method  and traditional methods  for the $^{17}$O$(\alpha,n)$ case. In the table, we show the mean number of events with standard deviation in parentheses, which are evaluated by executing the new method for 50 repetitions with different seeds. The values are rounded to the nearest integer.

As expected, the classification of GE in this case is less difficult than in the $^{17}$F$(\alpha,p)$ case, because of the larger difference in proton number between the heavy ion ($^{20}$Ne) and the beam particles ($^{17}$O). This can be seen by the relatively large diagonal elements and relatively small off-diagonal elements of the confusion matrix.

\begin{table}[!tbh]
    \centering
    \begin{tabular}{c|cc}
    $^{17}$O$(\alpha,n)$ & GE pres. meth.   & Not-GE pres. meth.  \\ \hline
    GE trad. meth. & 2860(7)   &  53(7)      \\
    Not-GE trad. meth. & 330(46)   &  81004(46)
    \end{tabular}
    \caption{Confusion matrix for the $^{17}$O$(\alpha,n)$ case showing the mean number of events with standard deviation in parentheses.
    }
    \label{tab:17_O}
\end{table}

\subsection{Sources of uncertainty} 
\label{s:uncert}
In Tables \ref{tab:17_F} and \ref{tab:17_O} the standard deviation  provides an estimate of how much uncertainty is incurred by the proposed approach in extracting golden events. There are multiple sources of this uncertainty in the four phases of the  approach. In phases 2 and 4, the models~(the neural network transformation and the classifier) comprise estimated parameters. Typically, these parameters are estimated by using repetitive updates~(using stochastic gradient descent) starting from a random initialization point. The stochastic nature of the repetitive updates contributes to parametric uncertainty~\cite{abadi2016tensorflow}. 

All machine learning models are plagued with the choice of training data that is used to estimate the parameters. Typically and in our methodology, the training data is uniformly sampled from a larger set, therefore contributing aleatoric uncertainty~\cite{abadi2016tensorflow}. Furthermore, the choice of thresholds in phase 1, 2, and 3 also contributes to uncertainty where different choices of thresholds governs the number of false positives~(number of not-golden events classified as golden) in the detection. In this work the parametric and aleatoric uncertainty of the proposed approach is summarized by the standard deviations obtained through repetition of the detection experiment.

\subsection{Other ML methods for GE detection~($^{17}$F$(\alpha,p)$)}
\label{s:otherMLmeth}
\begin{table}[!tbh]
    \centering
    \begin{tabular}{c|ccc}
    Methods          &   Total GE   & Correctly identified         &  Wrongly identified \\ \hline
     One-class SVM~\cite{scholkopf1999support}   &  51852  & 1540                    &  50312     \\
     Isolation Forest~\cite{cheng2019outlier}    &  53916  & 1194                    &  52722 \\
     Local Outlier Factor~\cite{cheng2019outlier}&  53766  & 1233                    &  52533 \\ 
     Elliptical Envelope~\cite{antonini2018smart}&  51782  & 776                     &  51006 \\ 
     Present                                     &  9793   &  1825                   &  7968     \\ 
    \end{tabular}
    \caption{Results with different outlier detection~(GE are the outliers to be detected) approaches for the $^{17}$F$(\alpha,p)$ data.
    }
    \label{tab:out}
\end{table}

To establish the efficiency of the new methodology, we compare our approach with several out-of-the-box outlier detection methods with the assumption that GE's are the outliers. We use one-class SVM~\cite{scholkopf1999support}, isolation forest~\cite{cheng2019outlier},  local outlier factor~\cite{cheng2019outlier}, and  elliptical envelope~\cite{antonini2018smart} for this purpose. First, we remark that all of these out-of-the-box methods are  computationally prohibitive for detecting GE from $10^9$ (1 billion) points. Therefore, we apply these methods to the result of phase 1 where most of the easy-to-identify background events have been removed. The first  four rows  in Table~\ref{tab:out} correspond to results of these outlier detection methods, and the last row corresponds to the methodology presented in this paper. 
In the table  we record the total number of events identified by all these methods as GE 
the total number of events that were correctly identified as GE, and  the total number of events that were wrongly identified. In this context, ``correctly'' or ``wrongly'' identified GE means that the GE agree or disagree with the ones found using the traditional method, respectively. We remark that, while each of the methods is able to correctly identify several GE's, the number of GE's identified by our approach is much greater than that identified by all the out-of-the-box methods. For example, our approach correctly detects a total of $1825$ GE's whereas one-class SVM detects only $1540$---which is the most number of events correctly detected by any out-of-the-box outlier detection method. While $1540$ is a reasonable number, the main drawback of one-class SVM and other outlier detection methods is the number of false positives. The one-class SVM misidentifies 50312 events as GE~(which is the minimum among  rows in Table~\ref{tab:out}). In contrast to these approaches, our approach wrongly identifies only $7968$ events as GE---a reduction by a factor of 7 in the number of false positives relative to the best anomaly detection approaches in the literature. The additional steps utilized in our approach not only improves the detection of GE but dramatically reduces the number of false positive---a strength of our methodology.

\section{Conclusion}
\label{s:conclusion}
We have developed a method for the strip-wise classification of specific $\alpha$-induced reactions taking place inside the MUSIC detector. The method consists of four phases where statistical and machine-learning methods are combined. 

We applied the new method to two experimental data sets, where the golden events (events of interest) correspond to $^{17}$F$(\alpha,p)^{20}$Ne and $^{17}$O$(\alpha,n)^{20}$Ne reactions. For most strips, the new method yields results such that the percent differences with respect to results found with traditional methods are within $\pm 20\%$.
In the strips closest to the ends of the detector, strip 1 and strip 13 or higher, we observe the strongest disagreements and thus conclude that the new method is not yet valid for those strips. 
That said, we found that our newly developed method outperforms several out-of-the-box methods for outlier detection. 

We expect that our method will serve as a foundation for the development of future AI/ML methods for the classification of MUSIC data with enhanced automation.

\bibliographystyle{elsarticle-num}
\bibliography{bib/physics, bib/aiml}
 
\end{document}